# Multiple photon detection and the localization of light energy

M. V. Lebedev
Institute of Solid State Physics,
Russian Academy of Sciences,
Chernogolovka

## Introduction

Modern theory of photon detection is based on the following three assumptions.
1. The probability of a single photon detection during an infinitesmall time interval $dt$ from a infinitesmall area $dS$ of the detector light sensitive surface is:

$$pdt = \eta I(t) dS dt \qquad (1)$$

   where $I(t)$ is the instantaneous intensity of light and $\eta$ the detector quantum efficiency.
2. Probability to detect two or more photons during such a small time interval from a such a small area is negligible.
3. The detection process introduces itself no correlations between photon detection events in two non-overlapping time intervals.

The intensity $I(t)$ can be regarded as a squared amplitude of a classical field or, more rigorous, as an operator in a quantum field theory. The consequence of this three assumptions is the well known Mandel's formula [1].

It will be shown in what follows that probabilities of photon detection can be found on the basis of consideration of the phase space sells filling for an ideal photon gas, which led Bose to the derivation of the Plank's law [2]. These probabilities are directly connected to photon localization properties consistent with the entropy of a dilute photon gas found by Einstein [3,4].

## Detection of classical and quantum particles

Thermal radiation in a closed cavity can be considered as a thermodynamic system with a great number of degrees of freedom staying in a thermal equilibrium with the cavity walls. According to general thermodynamic point of view, this radiation can be considered as a almost free one during the time interval small if compared with a characteristic interaction time between radiation and the cavity walls. This radiation field undergoes transitions from one of his states to other ones due to week interaction with a thermal bath so that its state vector in its phase space instantly moves between states with nearly maximal entropy in accordance with the given temperature $\Theta$. Let us denote ensemble averaging of possible field states in what follows with angle brackets and time averaging with overline. If one regards thermal radiation as an ideal gas of photons, filling the cavity volume $V$, one can get the Plank's law, as it was shown by Bose [2], dividing the phase space of this ideal gas into sells each with a volume of $\hbar^3$ and finding the entropy maximum of the system assuming different probabilities of sell occupation. The physical meaning of the phase space dividing in sells is connected directly with the uncertainty principle:

$$\Delta p_x \Delta p_y \Delta p_z \Delta x \Delta y \Delta z = \Delta p_x \Delta p_y \Delta p_z V = \hbar^3 \qquad (2)$$

For the gas of classical particles this elementary sell of a phase space can be made as small as wanted, because the uncertainty principle does not holds for classical particles. This difference between classical and quantum particles has important consequents for photon detection.

Indeed, the mean number of detected particles during a time interval $\tau$ for a flow of classical particles which are uniformly distributed in space with density $\rho$ and moving toward the detector with velocity $u$ will be simply the mean number of particles in a detection volume:

$$n_\tau = \rho u \tau S \qquad (3)$$

where $S$ is the detector sensitive area. In the limit $n_\tau \to 0$ there will be in most cases no particles in a detection volume and only sometimes one or even more particles. The quantity $n_\tau$ can be regarded in this limit as proportional to the probability to detect one particle. Particles are considered as independent ones, hence, the probability to detect two particles would be proportional to $n_\tau^2$ and can be neglected if $\tau \to 0$. But this consideration does not hold for quantum particles.

The phase space sell for quantum particles has no internal spatial-temporal structure. This means that in the case when the detection volume is reduced below the volume $V$ of the elementary sell in real space the probabilities to detect one, two or more particles should vanish proportionally to the detection volume, but the ratios between this probabilities should remain constant in contrast to the case of classical particles. This can be seen regarding a following experiment.

Let the cavity walls be ideally reflecting and cavity filled with a low density gas of two-level atoms. Consider a small hole in a cavity with a sensitive area of a photon detector placed in this hole. The detector is supposed to be thermally isolated from the cavity and kept at rather low temperature so that one can neglect its own thermal noise. Let the spectral response of the detector be sufficiently broad and centered at the resonance frequency of the two-level atoms filling the cavity. Consider the time interval of photon detection to be small enough so that the detection volume coincides with the cavity volume $V$. According to Bose consideration one should expect that the probabilities of photon registration should depend on the probabilities of phase space sells fillings and should be proportional to the number of sells which can be registered by the detector. This phase space sells filling probabilities remain unchanged if one reduces the cavity volume $V$ for example to $\dfrac{V}{2}$, because properties of thermal radiation does not depend on the volume of the cavity. The number of phase space sells on the other hand should be reduced in two times due to the appropriate increase of the sell volume in the momentum space. So one can see that relative probabilities of photon detection should remain unchanged after the reduction of the cavity volume accompanied with the appropriate reduction of the detection time interval.

## Localization of light energy and photon detection

The problem of photon detection can be directly connected with the localization of electromagnetic energy. Einstein showed that the relative probability to find a low density photon gas occupying the volume $v$ in the case it initially occupied the volume $v_0$ is [3]:

$$W = \left(\frac{v}{v_0}\right)^N \qquad (4)$$

where $N$ is the total number of photons. Let the volume $v$ differ from $v_0$ by a small value $\delta v$. We can therefore write:

$$W = \frac{Q_v}{Q_{v_0}} = \left(\frac{v}{v+\delta v}\right)^N \qquad (5)$$

where $Q_v$ and $Q_{v_0}$ denote probabilities for the photon gas to occupy the volumes $v$ and $v_0$ respectively. Then we have:

$$Q_{v_0} = Q_v\left(1 + \frac{\delta v}{v}\right)^N \approx Q_v\left(1 + N\frac{\delta v}{v}\right) \qquad (6)$$

because $\delta v$ is regarded as sufficiently small. The probability to find the gas occupying the volume $v_0$ consists thus of the probability to occupy the volume $v$ and the probability of filling the volume $\delta v$. This last volume can be regarded as the volume of the light sensitive region of the photon detector. The value

$$\frac{Q_d}{Q_v} = \frac{N}{v}\delta v \qquad (7)$$

will then represent the relative probability to detect some photons after opening the shutter isolating volume $v$ from $\delta v$. After ensemble averaging one gets:

$$\left\langle\frac{Q_d}{Q_v}\right\rangle = \left\langle\frac{N}{v}\right\rangle \delta v = Z\langle n\rangle \delta v \qquad (8)$$

where $Z$ is the number of the phase space sells and $\langle n\rangle$ - the average filling of a sell. Only two events are possible after opening of the shutter: a response of the detector available or no response of the detector, but detector response occurs at any nonzero number of photons entering the volume $\delta v$. For the time being one cannot say what the probabilities to detect one, two or more photons are, but we can state immediately from (8) that relative probabilities should depend on $\langle n\rangle$ only and should hence remain unchanged with altering $\delta v$.

# Derivation of the formula giving the instantaneous probabilities of photons detection

It's well known that Einstein succeeded in getting the Plank's law considering the elementary processes of light emission and absorption with two level atoms and introducing the hypothesis of stimulated emission [5]. In a thermal equilibrium holds:

$$\frac{\langle n \rangle}{\langle n \rangle + 1} = \frac{\langle N_2 \rangle}{\langle N_1 \rangle} \qquad (9)$$

where $\langle N_2 \rangle$ and $\langle N_1 \rangle$ are respectively the mean numbers of atoms in the excited state and in the ground state and $\langle n \rangle$ is the average filling of a phase space sell as determined by the Plank's law. We assume for simplicity that all the states of the two-level atoms are non-degenerate.

One can see that this expression holds for mean values only. Consider for example the high frequency case, when $\langle n \rangle \to 0$. In most cases there will be no photons in the cavity in this limit and only sometimes one or more photons could be found in a cavity. The instantaneous value of $\frac{n}{n+1}$ will exhibit strong fluctuations. If, for example, one photon is found in a phase space sell this proportion will be equal to $\frac{1}{2}$ but the ratio $\frac{N_2}{N_1}$ will always be small, because most atoms remain unexcited.

The quantity $\frac{\langle n \rangle}{\langle n \rangle + 1}$ determines the probabilities of phase space sells filling for the ideal photon gas:

$$P_n = \frac{1}{\langle n \rangle + 1} \left( \frac{\langle n \rangle}{\langle n \rangle + 1} \right)^n \qquad (10)$$

This expression is a geometrical probability distribution well known as a Bose-Einstein distribution. It can be derived from the consideration of a thermal equilibrium between a harmonic oscillator and a thermal bath. But such a distribution can arise also in a very general case if considering transitions in a quantum system under an external perturbation.

Consider a thermodynamic system which can absorb energy only in equal portions. If we denote the probability of one quantum absorption as $R_x$ and suppose that this probability does not depend on a number of absorbed quanta we get for the probabilities to absorb $k+1$ and $k$ quanta:

$$R_{k+1} = R_x R_k \qquad (11)$$

Or

$$R_k = R_0 R_x^k \qquad (12)$$

Summing up from $k = 0$ to infinity it is straightforward to obtain the probability of no absorption $R_0 = 1 - R_x$. $R_x$ is connected with the first momentum of the distribution as

$$R_x = \frac{\langle k \rangle}{\langle k \rangle + 1} \qquad (13)$$

In this consideration (12) gives the instantaneous probability to absorb $k$ quanta in the case the mean number of absorbed quanta is $\langle k \rangle$. It should be noted that (12) gives probabilities of $k$ quantum absorption in a single elementary process. Energy absorption means a momentum transfer also. This was emphasized by Einstein [6] while considering the momentum transfer in a spontaneous emission process. So we can apply (12) to absorption of quanta occupying the same sell only. To get the total absorption one have to multiply (12) by the number of equally populated phase space sells. It looks surprising: how can the system absorb from a phase space sell filled for example with two photons five or more ones? The only thing which could be said here is that absorption is not an energy measuring process but a coordinate measuring one. Measuring of the photon coordinate unavoidably leads to a great uncertainty in its momentum and hence in energy. Measuring of photon coordinate does not commute with measuring of its momentum, so when we try to measure the coordinate of the twice filled state with a given momentum we get fluctuating results, but the mean number of detected photons should be the sell filling, namely two.

Let us apply (12) to the problem of photons registration with a detector located in a small hole into a cavity wall as it was outlined above. When the detector is kept at low temperature we can assume that all atoms of the detector light sensitive volume are initially in a ground state. The number of these atoms is much more than 1 and the photon gas density is low $\langle n \rangle << 1$, so the absorption of one photon should have no effect on the probability of another absorption process. It is clear that the probabilities of photons detection should depend on the detection volume, that is, on the area of the light sensitive surface and time interval of the detection act. We can however suppose that the detection volume should enter the expressions as a coefficient just like it enters (8), so we will drop this coefficient in what follows. In accordance with previous discussion we postulate that the mean number of photons absorbed from a single phase space sell is just the filling of this sell. Then for a probability $U_k$ to detect $k$ photons we have to write:

$$U_k = \sum_{n=0}^{\infty} P_n \frac{1}{n+1} \left( \frac{n}{n+1} \right)^k \tag{14}$$

where $P_n$ are the probabilities of the phase space sells filling. One can see that the mean number of absorbed photons coincides with $\langle n \rangle$ in agreement with the Plank's law:

$$\sum_{n=0}^{\infty} P_n(\langle n \rangle) \frac{1}{n+1} \sum_{k=0}^{\infty} k \left( \frac{n}{n+1} \right)^k = \sum_{n=0}^{\infty} P_n(\langle n \rangle) n = \langle n \rangle \tag{15}$$

Moreover the mean value of the transition probability: $\langle R_x \rangle$ coincides with the mean transition probability $P_x = \frac{\langle n \rangle}{\langle n \rangle + 1}$:

$$\langle R_x \rangle = \left\langle \frac{n}{n+1} \right\rangle = \sum_{n=0}^{\infty} P_n \frac{n}{n+1} = 1 - \sum_{n=0}^{\infty} \frac{P_n}{n+1} = 1 - P_0 \sum_{n=0}^{\infty} \frac{P_x^n}{n+1} = 1 - \frac{P_0}{P_x} \sum_{n=0}^{\infty} \frac{P_x^{n+1}}{n+1} \tag{16}$$

The sum in (16) can be calculated:

$$\sum_{n=0}^{\infty} P_x^{n+1} \frac{1}{n+1} = \int \left( \frac{d}{dP_x} \sum_{n=0}^{\infty} P_x^{n+1} \frac{1}{n+1} \right) dP_x = \int \frac{1}{1-P_x} dP_x = -\ln(1-P_x) \approx P_x \tag{17}$$

Substituting this result in (16) one gets:
$$\left\langle \frac{n}{n+1} \right\rangle = 1 - P_0 = P_x = \frac{\langle n \rangle}{\langle n \rangle + 1} \tag{18}$$

In the limit $\langle n \rangle \to 0$ (14) reduces to only two terms in the sum over $n$: $U_0 = P_0$ and
$$U_k = P_1 \left(\frac{1}{2}\right)^{k+1}, \quad k \geq 1 \tag{19}$$

This result explains why one can detect more than one photon at very low light levels with considerable probability and gives the ability to understand experimental results [7-11].

## Photons detection during a finite time

Expression (14) gives the ensemble average. This corresponds to probabilities of photons detection in a single detector response. It is interesting to obtain probabilities of photons detection during a finite time interval $T$, when multiple detector responses are possible. One could expect that time averaging should tend to ensemble averaging with time interval tending to zero according to the ergodicity hypothesis. Let us call the event of some photons detection in a single detector response the elementary photo event. Expression (14) gives then the probabilities of elementary photo events. The probabilities of multiple elementary photo events can be found supposing the detection process to be stationary, that is the equivalence of all time moments. For a binary variable $w$, taking the value 0 with the probability $U_0 = P_0$ in the case of absence of the detector response and the value 1 with probability $1 - U_0 = P_x$ in the case of any detector response, the consideration usually made for obtaining the Mandel's formula obviously holds. So, for the probability distribution $G_m(T)$ to observe $m$ detector responses during the time interval $T$ one gets the Poisson distribution:
$$G_m(T) = \frac{\overline{m}^m}{m!} e^{-\overline{m}} \tag{20}$$

with $\overline{m} = P_x \frac{T}{\tau}$ and $\tau$ being the characteristic interaction time between radiation and the thermal reservoir. The probability $H_k$ to detect $k$ photons in a single detector response provided the detector response is available will be:
$$H_k = \frac{U_k}{1 - U_0} = \frac{1}{P_x} \sum_{n=0}^{\infty} P_n \frac{1}{n+1} \left(\frac{n}{n+1}\right)^k \qquad k \geq 1 \tag{21}$$

The detector response takes place if one of the alternative events happens, that's why:
$$\sum_{k=1}^{\infty} H_k = 1 \tag{22}$$

as can be checked easily. Individual detector responses are supposed to be statistically independent (assumption 3 of the modern theory of photon detection), that's why the probability space of elementary events for a finite time interval should be the product of probability spaces for individual responses.

$$\left(\sum_{k=1}^{\infty} H_k\right)^m = 1 \qquad m \geq 1 \qquad (23)$$

It is straightforward to obtain from (20) – (23) the probabilities to detect a given number of photons during a finite time interval $T$. One can see also that time averages tend to ensemble averages with $T \to \tau$.

## Conclusion

Thus we see that theory of photon detection of thermal light can be formulated without considering instantaneous light intensity but using Bose consideration of phase space sells filling for an ideal photon gas. For a detector unable to resolve the number of simultaneously absorbed photons we get just the Mandel's result (20) where instead of the instantaneous intensity stands the average transition probability $P_x$.